\documentclass[prd,aps,nofootinbib,twocolumn]{revtex4} 
\usepackage{amsmath,amssymb,amsfonts,array,eucal,mathrsfs,upgreek,stackrel,slashed,graphicx,tensor,color,enumerate}
\usepackage{hyperref}
\DeclareMathAlphabet{\mathpzc}{OT1}{pzc}{m}{it}
\newcommand{\ft}{\footnote}
\newcommand{\beqn}[1]{\begin{equation*}#1\end{equation*}}
\newcommand{\beql}[1]{\begin{equation}\label{#1}}
\newcommand{\eeq}{\end{equation}}
\newcommand{\eq}[1]{eq.~\eqref{#1}}
\newcommand{\lf}{\left}
\newcommand{\rf}{\right}
\newcommand{\fr}{\frac}
\newcommand{\rt}{\sqrt}
\newcommand{\tn}{\tensor}

\newcommand{\iim}{\Leftrightarrow}

\newcommand{\met}{\mathsf{g}}
\newcommand{\pmet}{\mathsf{p}}
\newcommand{\hatmet}{\hat{\mathsf{g}}}
\newcommand{\tlT}{\tilde{T}}
\newcommand{\tlN}{\tilde{N}}
\newcommand{\ed}{\mathrm{d}}
\newcommand{\dR}{\mathrm{d}R}
\newcommand{\dT}{\mathrm{d}T}
\newcommand{\LieD}{{\large\text{\pounds}}}
\newcommand{\dl}{\partial}
\newcommand{\Dl}{\nabla}
\newcommand{\vDl}{\vec{\nabla}}

\newcommand{\ie}{\textit{i.e.}~}

\newcommand{\Ric}{\mathpzc{R}}
\newcommand{\ric}{\mathscr{R}}
\newcommand{\EinG}{\mathpzc{G}}
\newcommand{\EEq}{\mathpzc{E}}
\newcommand{\EEqae}{\mathpzc{E}^{\ae}}
\newcommand{\EEqH}{\mathpzc{E}^{\textsc{h}}}
\newcommand{\ac}{\CMcal{S}}
\newcommand{\lag}{\mathscr{L}}

\newcommand{\aeT}{\mathpzc{T}^{\ae}}
\newcommand{\aeTH}{\mathpzc{T}^{\textsc{h}}}

\newcommand{\vAE}{\vec{\AE}}

\newcommand{\vN}{\vec{N}}

\newcommand{\fIM}{f_\textsc{im}}

\begin{document}
\title{Evolution and spherical collapse in Einstein-{\ae}ther theory and Ho\v{r}ava gravity}
\author{Jishnu Bhattacharyya,$^1$ Andrew Coates,$^1$ Mattia Colombo,$^1$ and Thomas P. Sotiriou$^{1, 2}$}
\affiliation{$^1$ School of Mathematical Sciences, University of Nottingham, University Park, Nottingham, NG7 2RD, United Kingdom\\
$^2$ School of Physics and Astronomy, University of Nottingham, University Park, Nottingham, NG7 2RD, United Kingdom} 
\date{\today}
\begin{abstract}
We compare the initial value formulation of the low-energy limit of (non-projectable) Ho\v{r}ava gravity to that of Einstein-{\ae}ther theory when the {\ae}ther is assumed to be hypersurface orthogonal at the level of the field equations. This comparison clearly highlights a crucial difference in the causal structure of the two theories at the non-perturbative level: in Ho\v{r}ava gravity evolution equations include an elliptic equation that is not a constraint relating initial data but needs to be imposed on each slice of the foliation. This feature is absent in Einstein-{\ae}ther theory. We discuss its physical significance in Ho\v{r}ava gravity. We also focus on spherical symmetry, and we revisit existing collapse simulations in Einstein-{\ae}ther theory. We argue that they have likely already uncovered the dynamical formation of a universal horizon and that they can act as evidence that this horizon is indeed a Cauchy horizon in Ho\v{r}ava gravity.
\end{abstract}
\maketitle
\section{Introduction}
Lorentz violating gravity theories have garnered interest in recent years. One key reason is that obtaining quantitative infrared constraints on Lorentz violation in the gravity sector requires a consistent parametrisation of deviations from Lorentz invariance in terms of a Lorentz-violating gravity theory~\cite{Jacobson:2000xp}. Additional recent motivation came from a concrete realisation of the idea that abandoning Lorentz symmetry can lead to improved ultraviolet behaviour in gravity~\cite{Horava:2009uw}.  

Lorentz symmetry is clearly intimately related with the causal structure of general relativity. The most extreme manifestation of this is the notion of a black hole, the existence of which one tends to associate with the behaviour of light rays in the vicinity of a horizon. Lorentz-violating gravity theories tend to include superluminal excitations -- in fact, constraints from vacuum \v{C}erenkov radiation seem to not leave much of an alternative~\cite{Elliott:2005va}. One then faces the question of whether a black hole can be said to exist in a meaningful way. And if so, what is the appropriate definition? The precise answer to these questions will actually depend on the theory under scrutiny. 

Einstein-{\ae}ther theory~\cite{Jacobson:2000xp} is a vector-tensor theory where the vector, referred to as the {\ae}ther, is constrained to be unit timelike. Because of this constraint the {\ae}ther cannot vanish even in flat spacetime and defines through its trajectories a preferred set of timelike curves that thread the spacetime. The theory propagates the usual spin-2 graviton as well as a spin-1 mode and a spin-0 mode. The propagation of these modes can be described by following null rays of separate metrics, $\tilde{\met}^{(i)}_{a b} = \met_{a b} - (s^2_i - 1)u_a u_b$, where $s_i$ is the speed of the spin-$i$ mode. Here $\met_{a b}$ is the metric that minimally couples to matter and whose null rays are photon trajectories. All modes generically have different propagation speeds  from each other and can be superluminal. A stationary black hole in this theory is then identified with a region cloaked by a succession of Killing horizons for $\met_{a b}$ and each of the metrics $\tilde{\met}^{(i)}_{a b}$~\cite{Eling:2006ec}. These horizons are referred to as spin-$i$ horizons.

Ho\v{r}ava gravity is a theory with a preferred foliation. The purpose of having a preferred foliation is to introduce higher order spatial derivatives that modify the propagator in the ultraviolet and render the theory power-counting renormalizable~\cite{Horava:2009uw}. It is precisely this feature of the theory that leads to dispersion relations of the type $\omega^2\propto k^6$ as $k\to \infty$. This seems to present a challenge to defining black holes as there is no upper limit for the speed of perturbations. A less evident but even more worrisome feature when it comes to black holes is that, even at the low-energy limit where the dispersion relations become linear, there is still instantaneous propagation, in the form of an elliptic mode~\cite{Blas:2011ni}. 

Perhaps surprisingly, black holes can still be defined in a meaningful way. The boundary of the causally disconnected region is called a `universal horizon'~\cite{Barausse:2011pu,Blas:2011ni}, owing to the fact that no signals of any speed can leave it. In terms of the preferred foliation this horizon corresponds to a leaf which is disconnected from spatial infinity~\cite{Bhattacharyya:2015gwa}. Since propagation into the future means crossing the leaves of the foliation in a given direction, if such a leaf exists no signal emitted inside it can ever cross it toward spatial infinity without travelling toward the past. Universal horizons were initially found in static, spherically symmetric black holes~\cite{Barausse:2011pu,Blas:2011ni,Berglund:2012bu} but they seem to be a more generic feature, as they have also been shown to exist in slowly rotating black holes~\cite{Barausse:2012qh} and in lower-dimensional rotating black holes~\cite{Sotiriou:2014gna}.

Because of the close relation between Einstein-{\ae}ther theory and the low energy limit of Ho\v{r}ava gravity~\cite{Jacobson:2010mx}, which we will discuss thoroughly in the next section, the known static, spherically symmetric, asymptotically flat solutions of the former are also solutions of the latter and vice versa~\cite{Barausse:2012ny}. This is no longer true for slowly-rotating solutions~\cite{Barausse:2012ny,Barausse:2012qh}. Moreover, it is not known if non-static, spherically symmetric solutions in the two theories match, so it is not clear if spherical collapse in the two theories is identical or simply leads to the same end state.

This is one of the questions we are seeking to answer. More generally, we will perform a comparison of the initial value formulation of Ho\v{r}ava gravity and that of Einstein-{\ae}ther theory, under the additional assumption that the {\ae}ther in the latter is hypersurface orthogonal at the level of the field equations. Under this assumption the {\ae}ther defines a foliation (as opposed to just a preferred frame) and this makes the two theories resemble each other. Hence, making this assumption suppresses some obvious differences, such as the existence of propagating spin-1 modes in Einstein-{\ae}ther theory, and allows one to focus on some more subtle ones, such as the different standings of the preferred foliation in each theory and the dependence of evolution on (future) boundary data in Ho\v{r}ava gravity.

We will also consider the special case of spherical symmetry in more detail. Spherical collapse in Einstein-{\ae}ther theory has been considered in Ref.~\cite{Garfinkle:2007bk}. To date there are no collapse simulations in Ho\v{r}ava gravity (see, however, Ref.~\cite{Saravani:2013kva}). Moreover, Ref.~\cite{Garfinkle:2007bk} predates the introduction of the notion of a universal horizon, which was later understood to be one of the most prominent features of spherical black holes in both Einstein-{\ae}ther and Ho\v{r}ava gravity~\cite{Barausse:2011pu}. Clearly, whether universal horizons actually form from collapse is a key question. Additionally, it is important to verify that the known static solutions that harbour them are indeed the endpoints of gravitational collapse in Lorentz violating theories. We will revisit the results of Ref.~\cite{Garfinkle:2007bk} and attempt to understand whether dynamical formation of universal horizons had indeed been found. We will also discuss possible improvements in future simulations that could answer this question unambiguously. Our analysis will also shed light on whether these simulations can be seen  as modelling collapse in Ho\v{r}ava gravity as well. 

\section{The theories}
\label{theories}

By introducing a Stuckelberg field, Ho\v{r}ava gravity can be reformulated in a manifestly covariant manner~\cite{Jacobson:2010mx} as a scalar-tensor theory, where the tensor degree of freedom is the metric $\met_{a b}$ as usual, while the `non-metric' degree of freedom is a scalar field $T$. The level surfaces of this scalar are the leaves of the preferred foliation; hence $T$ is constrained to have a~\emph{timelike gradient everywhere}. Therefore, one may introduce a timelike unit one-form $u_a$ such that
	\beql{HSO:ae}
	u_a = -N\Dl_a T~,
	\eeq
by virtue of being  orthogonal to the constant $T$ hypersurfaces, and
	\beql{ae:norm}
	u_a u_a \met^{a b} = u\cdot u=-1~,
	\eeq
on account of being unit timelike. Combining the two conditions yields
	\beql{solve:N}
	N^{-2} = -\met^{a b}(\Dl_a T)(\Dl_b T)~.
	\eeq
In accordance with common practice, we will call the normalized gradient of $T$, \ie $u_a$, the~\emph{{\ae}ther}.

$T$ plays the role of a preferred time in Ho\v{r}ava theory, but it should be noted that the theory is invariant under reparametrizations of the type
	\beql{T:reparam}
	T \mapsto \tlT = \tlT(T)~,
	\eeq
where $\tlT(T)$ is some arbitrary function of $T$. That is, $T$ defines an ordered  preferred slicing but does not introduce a preferred labelling of the slices.  Under such reparametrizations, the function $N$ is required to transform as
	\beql{N:reparam}
	N \mapsto \tlN = (\ed\tlT/\dT)^{-1}N~,
	\eeq
in accordance with~\eq{solve:N} such that $u_a$ as well as quantities built out of $u_a$ and its covariant derivatives are invariant under the above reparametrizations.

Consider the action
\beql{ac:HL}
\ac = \fr{1}{16\pi G}\int\ed^4x\,\rt{-\met}\lf(\Ric + \lag \rf)~,
	\eeq
where $G$ is a coupling constant with suitable dimensions, $\Ric$ is the four-dimensional curvature scalar, and $\lag$ is given in terms of the derivatives of the {\ae}ther as
	\beql{lag:HL}
	\lag= -Z^{a b c d}(\Dl_a u_c)(\Dl_b u_d)~,
	\eeq
so that everything is manifestly invariant under the reparametrizations~\eqref{T:reparam}. The tensor $Z^{a b c d}$ defining the Lagrangian~\eqref{lag:HL} is given by
	\beql{def:Zabcd}
	\begin{split}
	Z^{a b c d} = c_1\met^{a b}\met^{c d} + c_2\met^{a c}\met^{b d} & + c_3\met^{a d}\met^{b c} \\
	& - c_4 u_e u_f \met^{a e}\met^{b f}\met^{c d}~,
	\end{split}
	\eeq
with coupling constants $c_1, \dots, c_4$  allowing for the most general two-derivative action for a unit one-form field. Adopting $T$ as a time coordinate, on the premises that its gradient is always timelike, introduces a foliation defined by the constant-$T$ surfaces. Action (\ref{ac:HL}) then becomes the second derivative truncation of Ho\v{r}ava gravity~\cite{Jacobson:2010mx}, {\em i.e.}~the low-energy part of the theory. Once one has adopted $T$ as a time coordinate, the residual symmetry  is invariance under diffeomorphisms that preserve the foliation. Indeed, the full action of Ho\v{r}ava gravity includes all the terms that respect this symmetry and contain up to sixth order spatial derivatives in the preferred foliations~\cite{Blas:2009qj}.\footnote{We will not consider here versions of the theory where extra restrictions have been imposed, such as projectability or detailed balance~\cite{Horava:2009uw,Sotiriou:2009gy,Sotiriou:2009bx,Weinfurtner:2010hz,Mukohyama:2010xz,Vernieri:2011aa,Vernieri:2012ms,Sotiriou:2010wn}.}

Here we will only consider the low-energy part of Ho\v{r}ava gravity so we will not discuss these higher-order terms any further. However, some remarks are in order. These terms are higher-order in spatial derivatives, and they do not contain any time derivatives. This underscores the existence of a preferred foliation in Ho\v{r}ava gravity. Even though these terms can be written in a manifestly covariant way in the same fashion as the low-energy part of the action, in such a covariant formulation the full theory would appear highly fine-tuned (as higher-order time derivatives would have to cancel out)~\cite{Sotiriou:2011dr}. Moreover, discarding the higher-order terms does not mean that the preferred foliation ceases to be preferred. As we noted above, even in the low-energy theory that can be described in a covariant manner by action (\ref{ac:HL}), $T$ has to be nonzero and have a timelike gradient in {\em every} solution, thereby signifying that every solution comes with a special foliation. Additionally, action (\ref{ac:HL}) actually contains more than two derivatives of $T$, which is an indication that the theory will not satisfy second order differential equations in a generic foliation. 

Indeed, a variation of action~\eqref{ac:HL} (up to boundary terms) gives 
	\beql{de-ac:HL}
\delta\ac = \fr{1}{16\pi G}\int\ed^4x\,\rt{-\met}\lf[\EEqH_{a b}\,\delta\met^{a b} + 2(\Dl_a[N\vAE^a])\delta T\rf]~,
	\eeq
where Eq.~\eqref{HSO:ae} has been taken into account (recall that $T$ and not the \ae ther is the fundamental field here).
The tensor $\EEqH_{a b}$ is defined as
	\beql{def:EEqH}	
	\EEqH_{a b} = \EinG_{a b} - \aeTH_{a b}~,
	\eeq
where $\EinG_{a b}$ is the four-dimensional Einstein tensor and $\aeTH_{a b}$ is $T$'s stress-energy tensor. $\AE^a$ is the functional derivative of the {\ae}ther Lagrangian~\eq{lag:HL} with respect to the {\ae}ther, 
	\beql{def:vAE}
	\vAE^a \equiv \tn{\pmet}{^a_b}\AE^b~,
	\eeq
and $\pmet_{a b}$ is the projector onto the constant khronon leaves
	\beql{def:pmet}
	\pmet_{a b} = \met_{a b} + u_a u_b~,
	\eeq
also acting as the induced metric on the leaves of the preferred foliation. Therefore, $\vAE^a$ is manifestly orthogonal to the {\ae}ther by construction. From~\eq{de-ac:HL}, the equations of motion of Ho\v{r}ava gravity are then
	\beql{HL:EOM}
	\EEqH_{a b} = 0~, \qquad \Dl_a[N\vAE^a] = 0~.
	\eeq
$\vAE^a$ already contains the second derivative of the \ae ther, which implies that Eq.~(\ref{HL:EOM}) contains third order time derivatives in an arbitrary foliation. However, the fact that $\vAE^a$ is orthogonal to the {\ae}ther implies that (only) in the preferred foliation defined by $T$, the divergence in Eq.~(\ref{HL:EOM}) is purely spatial and there are only two time derivatives~\cite{Jacobson:2010mx}.

The other theory we will consider, namely Einstein-{\ae}ther theory~\cite{Jacobson:2000xp}, is a true vector-tensor theory.  The fundamental fields are the metric and the {\ae}ther. The \ae ther was treated as a vector in the original formulation~\cite{Jacobson:2000xp} of the theory, but treating it as a one-form ({\em i.e.}~keeping the variation of $u_a$ fixed on the boundary) leads to the same theory (see also~\cite{Barausse:2012qh} for a discussion). The equations of motion of this theory can be derived from an action that is \emph{formally identical} to~\eq{lag:HL}, but the {\ae}ther is constrained to satisfy \emph{only} the unit norm constraint of \eq{ae:norm} and is not hypersurface orthogonal in general. Variation with respect to the metric and the {\ae}ther yields 
	\beql{de-ac:AE}
\delta\ac= \fr{1}{16\pi G}\int\ed^4x\,\rt{-\met}\lf[\EEqae_{a b}\,\delta\met^{a b} + 2\vAE^a\,\delta u_a\rf]~,
	\eeq
where the unit constraint of Eq.~\eqref{ae:norm} has been imposed by constraining the {\ae}ther's  variation. $\EEqae_{a b}$ is defined as
	\beql{def:EEqae}
	\EEqae_{a b} = \EinG_{a b} - \aeT_{a b}~,
	\eeq
with $\aeT_{a b}$ being the stress energy tensor of the {\ae}ther. Thus from~\eq{de-ac:AE}, the equations of motion of Einstein-{\ae}ther theory are
	\beql{ae:EOM}
	\EEqae_{a b} = 0~, \qquad \vAE^a = 0~.
	\eeq

$\aeT_{a b}$ and $\aeTH_{a b}$ are formally identical as they come from formally identical actions under variation with respect to the metric. This means that, if one imposes the hypersurface orthogonality condition~\eqref{HSO:ae} on the {\ae}ther as an additional simplifying assumption at the level of the equations of motion in Einstein-{\ae}ther theory, then the systems of equations \eqref{HL:EOM} and \eqref{ae:EOM} will have the same `Einstein equations.' Moreover, any such hypersurface-orthogonal solution of Einstein-{\ae}ther theory will also be a solution of Ho\v{r}ava gravity. The converse is not generically true. It has, however, been shown to hold for spherically symmetric, asymptotically flat solutions under the assumption that all leaves of the foliation reach the centre and the centre is regular~\cite{Blas:2010hb}. It has also been shown to hold for static, spherically symmetric, asymptotically flat solutions without any further assumptions~\cite{Barausse:2012ny}, as well as for static, spherically symmetric solutions with more general asymptotics but with a regular universal horizon~\cite{Bhattacharyya:2014kta}.

\section{Evolution}
\label{evo}

As a genuine vector-tensor theory, Einstein-{\ae}ther theory propagates not only the standard spin-2 mode of general relativity but also spin-1 and spin-0 modes. In Ho\v{r}ava gravity, on the other hand, $u_a$ is hypersurface orthogonal and given in terms of $T$  via~\eqref{HSO:ae}, so there cannot be a dynamical spin-1 mode. This is certainly a significant difference in the dynamics of the two theories.

However, here we wish to focus on the more subtle differences that are not related to the existence of vector modes. To this end, we wish to compare the (non-perturbative) dynamics of Ho\v{r}ava gravity with that of Einstein-{\ae}ther theory when the {\ae}ther is constrained to be hypersurface orthogonal  (at the level of the equations) throughout the evolution.\footnote{Note that, in general, evolution could generate vorticity, so our condition is stronger that selecting vorticity-free initial data. It might well be that constraining the aether to be hypersurface orthogonal throughout the evolution could lead to an overconstrained system in the absence of extra symmetries, {\em e.g.}~spherical symmetry. This is an interesting open question, but it will not concern us here, as we simply seek for the most general setting in which we can straightforwardly compare Einstein-{\ae}ther theory and Ho\v{r}ava gravity.}  An important subcase that we will discuss later in more detail will be spherically symmetric collapse, as spherically symmetric vectors are hypersurfance orthogonal. However, in this section we will opt to be as general as possible, and we will not assume any symmetries.

By virtue of Frobenius's theorem~\cite{Wald:1984rg}, a hypersurface orthogonal (hence twist-free) unit timelike {\ae}ther satisfies
	\beql{HSO:ae:Frobenius}
	\Dl_a u_b - \Dl_b u_a = -u_a a_b + u_b a_a~,
	\eeq
where $a_a$ is the acceleration of the {\ae}ther congruence defined as usual
	\beql{def:acc}
	a^a = \Dl_u u^a \qquad\iim\qquad a_a = N^{-1}\vDl_a N~.
	\eeq
Here $\vDl_a$ is the projected covariant derivative on the preferred foliation, and the second expression in~\eq{def:acc} above is a consequence of Eqs.~\eqref{HSO:ae} and~\eqref{HSO:ae:Frobenius}. One may thus expand the covariant derivative of the {\ae}ther as
	\beql{Du}
	\Dl_a u_b = -u_a a_b + K_{a b}~,
	\eeq
where $K_{a b}$ is the extrinsic curvature of the preferred leaves due to their embedding in spacetime,
	\beql{def:Kab}
	K_{a b} = \fr{1}{2}\LieD_{u}\pmet_{a b}~,
	\eeq
and is purely spatial (\ie orthogonal to the {\ae}ther) by definition. The mean curvature $K$, \ie the trace of the extrinsic curvature tensor, is then given by
	\beql{def:trK}
	K = \met^{a b}K_{a b} = \pmet^{a b}K_{a b} = (\Dl\cdot u)~.
	\eeq
We may now use the above quantities and relations to adapt the equations of motion of both theories, Eqs.~\eqref{HL:EOM} and \eqref{ae:EOM}, to the foliation defined by the {\ae}ther. For Ho\v{r}ava gravity this is imperative as mentioned earlier, for it is only in this foliation that the equations become second-order in time derivatives. This is simply the preferred foliation determined by $T$, in which the theory is usually defined. For Einstein-{\ae}ther theory, however, this is simply a choice which we make in order to facilitate the comparison with Ho\v{r}ava gravity.  It is also worth noting that, even though we are adopting a foliation, we will refrain from adopting any coordinate system. 

As already noted, for a hypersurface orthogonal {\ae}ther Einstein's equations in both the theories are formally identical. One may furthermore show~\cite{Barausse:2011pu, Jacobson:2011cc} that in that case, the covariantized Bianchi identities are formally identical as well. When adapted to the preferred foliation, the (generalised) Bianchi identities for both the theories read
	\beql{Bianchi}
	\begin{split}
	                                            \dl_T\tn{\EEq}{^T_i} & = 0~, \\
	\dl_T\tn{\EEq}{^T_T} + (\rt{-\met})^{-1}\dl_i[\rt{-\met}N\vAE^i] & = 0~,
	\end{split}
	\eeq
where $i = \{1, 2, 3\}$ denote~\emph{coordinate indices} on the preferred leaves, $\EEq_{a b}$ is either of $\EEqH_{a b}$~\eqref{def:EEqH} or $\EEqae_{a b}$~\eqref{def:EEqae} (for a hypersurface orthogonal {\ae}ther), and in writing Eq.~\eqref{Bianchi} it was assumed that all Einstein's equations (but not the {\ae}ther/$T$ equations) are satisfied~\emph{on the given leaf}. Note that a $(1 + 3)$ decomposition of the equations of motion of Einstein-{\ae}ther theory need not necessarily be performed with respect to the {\ae}ther's foliation, and in general, the corresponding constraint equations are a combination of Einstein's equations and the {\ae}ther's equation of motion. However, (only) when formulated as a theory of a one-form, the constraint equations of Einstein-{\ae}ther theory adapted to the {\ae}ther's foliation do not involve the {\ae}ther's equations of motion but only the Einstein's equations in the  form of $\tn{(\EEqae)}{^T_T} = \tn{(\EEqae)}{^T_i} = 0$. Thus according to Eq.~\eqref{Bianchi}, these constraints are also preserved in time once the {\ae}ther becomes `on shell'~\eqref{ae:EOM} as well. For Ho\v{r}ava gravity, on the other hand, the only allowed foliation to perform a $(1 + 3)$ decomposition of the equations of motion with respect to~\emph{is} the preferred foliation; only then proper constraint equations can be found in the form of $\tn{(\EEqH)}{^T_T} = \tn{(\EEqH)}{^T_i} = 0$ which are first order in the $T$-derivative and according to Eq.~\eqref{Bianchi} are preserved in time once the khronon becomes `on-shell'~\eqref{HL:EOM}.

In both theories, the constraint $\tn{\EEq}{^T_T} = 0$ constitutes the~\emph{energy constraint equation} explicitly given by
	\beql{Ceqn:E}
	(1 - c_{13})K_{a b}K^{a b} - (1 + c_2)K^2 + c_{14}\lf[2(\vDl\cdot a) + a^2\rf] - \ric = 0~,
	\eeq
where $\ric$ is the intrinsic scalar curvature of the leaves of the foliation. The equations $\tn{\EEq}{^T_i} = 0$ become the~\emph{momentum constraint equations} and take the form
	\beql{Ceqn:p}
	(1 - c_{13})\vDl_c\tn{K}{^c_a} = (1 + c_2)\vDl_a K~.
	\eeq
The remaining Einstein's equations (\ie those completely projected onto the preferred foliation), for both theories, reduce to an~\emph{evolution equation} for the mean curvature
	\beql{EEq:Ruu}
	\begin{split}
	c_{\ell}\Dl_uK = & - (1 - c_{13})K_{a b}K^{a b} - \fr{c_{123}}{2}K^2 \\
	                 & + \lf[1 - \fr{c_{14}}{2}\rf](\vDl\cdot a + a^2)~,
	\end{split}
	\eeq
where $c_{123} = c_2 + c_{13}$ and $c_{\ell} = 1 + (1/2)c_{13} + (3/2)c_2$, and an~\emph{evolution equation} for the traceless part $[K]_{a b} = K_{a b} - (K/3)\pmet_{a b}$
	\beql{EEq:ij}
	\begin{split}
	& \LieD_{u}[K]_{a b} = 2[K]_{a c}\tn{[K]}{_b^c} - \fr{K}{3}[K]_{a b} \\
	& + \fr{1 - c_{14}}{1 - c_{13}}\lf[a_a a_b - \fr{a^2\pmet_{a b}}{3}\rf] \\
	&  + \fr{1}{1 - c_{13}}\lf[\vDl_a a_b - \fr{(\vDl\cdot a)\pmet_{a b}}{3} - \ric_{a b} + \fr{\ric\pmet_{a b}}{3}\rf],
	\end{split}
	\eeq
where $\ric_{a b}$ is the Ricci curvature of the preferred hypersurfaces. Collectively, Eqs.~\eqref{Ceqn:E},~\eqref{Ceqn:p},~\eqref{EEq:Ruu}, and~\eqref{EEq:ij} provide all Einstein's equations for both the theories.

As noted above, Eqs.~\eqref{EEq:Ruu} and~\eqref{EEq:ij} provide a set of~\emph{evolution equations for the extrinsic curvature} that are first order in time derivatives with respect to the preferred foliation in both the theories (for Einstein-{\ae}ther theory, these equations were already obtained in~\cite{Garfinkle:2007bk}). Taken together with~\eq{def:Kab}, these provide a set of first order evolution equations for the pair of conjugate variables consisting of the components of the induced metric and those of the extrinsic curvature. To turn these equations explicitly into a set of coupled partial differential equations, one needs to introduce a set of coordinates on the leaves of the hypersurfaces and perform a lapse-shift decomposition of the metric. However, this goes beyond our current goal; we merely wish to point out that even in the most general setting the `metric-extrinsic curvature pair' can be evolved in the same manner with respect to the preferred foliation in both the theories. The difference between the dynamics of the theories --- and the related issue of the existence of the instanteneous mode in Ho\v{r}ava theory --- stems from the evolution of the {\ae}ther/$T$. We will take this issue up next and study it in detail in spherical symmetry. 
\section{The instantaneous mode of Ho\v{r}ava gravity}\label{instant}
In terms of the kinematic variables introduced previously, the quantity $\vAE_a$~\eqref{def:vAE} is given by
	\beql{vAE:HSO}
	\vAE_a = \fr{c_{123}}{(1 - c_{13})}\vDl_a K - c_{14}[Ka_a + \LieD_{u} a_a - 2K_{a c}a^c]~.
	\eeq
This allows one to interprete the {\ae}ther's equation of motion~\eqref{ae:EOM} in Einstein-{\ae}ther theory as an evolution equation for the acceleration as follows~\cite{Garfinkle:2007bk}
	\beql{ae:EOM:ae}
	\LieD_{u} a_a = 2K_{a c}a^c - Ka_a + \fr{c_{123}}{c_{14}(1 - c_{13})}\vDl_a K~.
	\eeq
Needless to say, the above is not satisfied, in general, in Ho\v{r}ava gravity; instead, the `khronon's equation of motion' in Ho\v{r}ava gravity is given by~\eqref{HL:EOM}
	\beql{ae:EOM:HL}
	\Dl_a[N\vAE^a] = 0 \qquad\iim\qquad \vDl_a[N^2\vAE^a] = 0~,
	\eeq
where the expression for $\vAE^a$ is identical with that given in~\eq{vAE:HSO}. The difference in the dynamics in the two theories thus lies in the difference between the nature of Eqs.~\eqref{ae:EOM:ae} and~\eqref{ae:EOM:HL}. To study them closely, we will henceforth restrict ourselves to spherical symmetry, which will allow us to integrate~\eqref{ae:EOM:HL} very easily. Note that in spherical symmetry, the hypersurface orthogonality of the {\ae}ther is guaranteed kinematically.

Toward setting up a suitable coordinate system that makes the spherical symmetry manifest (among other things), let us start with some basic observations: in any coordinate system adapted to the {\ae}ther's foliation, the time coordinate is identical to $T$ [and hence subject to the reparametrizations~\eqref{T:reparam}]. Next, the unit spacelike vector $s_a$ along the acceleration,
	\beql{def:s_a}
	a_a = (a\cdot s)s_a~, \qquad (s\cdot s) = 1~, \qquad (u\cdot s) = 0~,
	\eeq
defines a natural spacelike direction in the spacetime which is orthogonal to the spherical directions by virtue of spherical symmetry. In order to be completely general (and in particular, to make our subsequent conclusions manifestly independent of any `gauge choices') we will now introduce a coordinate system adapted to the preferred foliation consisting of the `time coordinate' $T$ and a `radial coordinate' $R$ in which [along with~\eq{HSO:ae}]
	\beql{def:R}
	\begin{split}
	u^a & = N^{-1}\dl_T - N^{-1}N^R\dl_R~, \\
	s^a & = S^{-1}\dl_R~, \\
	s_a & = SN^R\Dl_a T + S\Dl_a R~,
	\end{split}
	\eeq
such that the functions $N = N(T, R)$, $S = S(T, R)$, and $N^R = N^R(T, R)$ describe the {\ae}ther configuration completely in a manifestly spherically symmetric manner. Note that for the above choice of coordinates, the shift vector is $\vN^a = N^R\dl_R$. Furthermore, the projector~\eqref{def:pmet} can be written as
	\beql{def:hatmet}
	\pmet_{a b} = s_a s_b + \hatmet_{a b}~,
	\eeq
where $\hatmet_{a b}$ is the metric on a unit two-sphere up to a conformal factor which is the~\emph{areal radius} $r$ squared
	\beqn{
	r = \rt{\fr{\text{Area of two-sphere}}{4\pi}}~, \qquad \hatmet_{a b} = r^2(\ed\theta^2 + \sin^2\theta\ed\phi^2)~,
	}
and $\theta$ and $\phi$ are the usual polar coordinates on the unit two-sphere. In what follows, $r$ will not be treated as a coordinate. Rather, the coordinate system that we have constructed consists of the coordinate functions $\{T, R, \theta, \phi\}$, and the areal radius is given as $r = r(T, R)$, in the same way as $N$, $S$, and $N^R$. Thus in the present coordinate system the full metric is
	\beql{met:TR}
	\met_{a b} = -N^2\dT^2 + S^2(N^R\dT + \dR)^2 + r^2(\ed\theta^2 + \sin^2\theta\ed\phi^2)~,
	\eeq
and the {\ae}ther and metric are completely specified by the four functions $N(T, R)$, $N^R(T, R)$, $S(T, R)$ and $r(T, R)$. Hence, the equations of motion of the two theories along with some suitable gauge choice will allow us to solve for these functions.

We may now integrate $T$'s equation of motion in Ho\v{r}ava gravity~\eqref{ae:EOM:HL} to obtain
	\beql{s.AE:ss}
	\dl_R\lf[r^2SN^2\vAE^R\rf] = 0 \qquad\iim\qquad (s\cdot\vAE) = \fr{\fIM(T)}{r^2N^2}~,
	\eeq
where $\fIM(T)$ is a `constant' of integration. Plugging this into the expression~\eqref{vAE:HSO} of $\vAE_a$ we then end up with a first order evolution equation for the acceleration very similar to~\eqref{ae:EOM:ae}
	\beql{dl_T:a.s}
	\dl_T(a\cdot s) = N^R\dl_R(a\cdot s) - N\hat{K}(a\cdot s) + \fr{c_{123}N\dl_R K}{c_{14}(1 - c_{13})S} - \fr{\fIM(T)}{c_{14}r^2N},
	\eeq
where $\hat{K} = \hatmet^{a b}K_{a b}$. This equation contains, in the most explicit manner, the most crucial difference between the dynamics of Einstein-{\ae}ther and Ho\v{r}ava theories. Indeed, one obtains the {\ae}ther's equation of motion in Einstein-{\ae}ther theory~\eqref{ae:EOM:ae} upon setting $\fIM(T) = 0$~\emph{for all} $T$, while $\fIM(T) \neq 0$ characterizes those solutions of Ho\v{r}ava theory which are~\emph{not} solutions of Einstein-{\ae}ther theory. Finally, as soon as one solves for $(a\cdot s)$ via~\eqref{dl_T:a.s} one may solve for the lapse by integrating~\eqref{def:acc} on a given $T$ slice, \ie
	\beql{a->N}
	\dl_R\log N = (a\cdot s)S~,
	\eeq
which implies
	\beql{int:a->N}
	\log N(T, R) = \log N(T, \infty) + \int_{\infty}^{R}\dR'(a\cdot s)S(T, R').
	\eeq
In this manner, all the relevant functions determining the spacetime-{\ae}ther/$T$ configuration in both the theories can be solved for.

In both theories there is still the reparametrization freedom of~\eq{T:reparam}. In Ho\v{r}ava gravity, this is a symmetry of the theory itself, whereas in Einstein-{\ae}ther theory it comes as a consequence of our restriction that the {\ae}ther be hypersurface orthogonal. Equation \eqref{N:reparam} implies that $\log N$ picks up a function of $T$~\emph{additively} under the reparametrization \eqref{T:reparam}. We can choose to reparametrize $T$ such that 
	\beql{N:bc}
	\log N(T, \infty) = 0
	\eeq
(which was also the choice in Ref.~\cite{Garfinkle:2007bk}). It then becomes apparent that generic Ho\v{r}ava gravity solutions are characterized by a non-zero $\fIM(T)$ while $\fIM(T) = 0$ in Einstein-{\ae}ther theory.

The root of the difference between the two theories is the following: turning the $T$ equation into an evolution equation for the lapse $N$ in Ho\v{r}ava gravity involves integrating a divergence on each slice.\footnote{Recall that we are working in the preferred foliation, so the lapse $N$ cannot be set to a constant by making a gauge choice, and hence it should be determined by the field equations.} Hence there is an elliptic part in this system of equations that is absent in Einstein-{\ae}ther theory. It should be stressed that this elliptic part is fundamentally different from the constraint equations, even though the latter are also elliptic. The main difference has to do with the fact that constraints are preserved by time evolution and hence need to be imposed only on an initial slice, while the divergence in Eq.~\eqref{s.AE:ss} has to be integrated on {\em every slice} and $\fIM(T)$ is to be determined by suitable boundary/asymptotic conditions. This will be discussed in more detail below. For instance, for generic functional forms of $\fIM(T)$,~\eq{dl_T:a.s} is singular if either $r = 0$ or $N = 0$. Thus, the physical requirement of regularity at the centre or on a universal horizon where $N=0$ for our choice of $T$ can impose $\fIM(T)=0$.

This is simply the non-perturbative manifestation of the instantaneous mode discussed in Ref.~\cite{Blas:2011ni} in  a perturbative setting. Indeed, when spherical perturbations around a black hole were considered in Ref.~\cite{Blas:2011ni}, it was the assumption of regularity on the universal horizon that forced the instantaneous mode to vanish, in agreement with what has been mentioned above.  

The above conclusions can be generalized beyond spherical symmetry, albeit somewhat qualitatively. To that end, we may begin by recalling that in diffeomorphism invariant scalar-tensor theories the equation determining the scalar field is dynamically redundant, as it can be obtained by taking a divergence of the field equations for the metric (see the Appendix). Hence, one can in principle solve the latter only and neglect the scalar's equation altogether. Since Ho\v{r}ava gravity can be written as a diffeomorphism invariant scalar-tensor theory, one can apply this logic. This then implies that consistent solutions can be obtained by solving only Eqs.~\eqref{Ceqn:E}-\eqref{EEq:ij} (where we have conveniently neglected Eq.~\eqref{HL:EOM} only after forming the constraint equations). Equation \eqref{Ceqn:E} can then be turned into the following Poisson type elliptic equation for $\varrho$, defined through $N = \varrho^2$:
	\beql{Ceqn:E:rho}
	\vDl^2\varrho = \fr{\varrho}{4c_{14}}\lf[\ric - (1 - c_{13})K_{a b}K^{a b} + (1 + c_2)K^2\rf].
	\eeq
As already pointed out in Ref.~\cite{Donnelly:2011df}, this  equation allows one to solve for the lapse $N$ {\em on each slice of the preferred foliation}. One can then subsequently compute the acceleration from~\eq{def:acc}. Thus, Eqs.~\eqref{Ceqn:E:rho},~\eqref{Ceqn:p},~\eqref{EEq:Ruu}, and~\eqref{EEq:ij} provide a complete set of equations that can dynamically determine the spacetime and the foliation in Ho\v{r}ava gravity. 

Since~\eq{Ceqn:E:rho} is a second-order~\emph{elliptic} equation in $\varrho$ that is not preserved by time evolution when $T$ is not taken to be on-shell [c.f.~Eqs.~\eqref{Bianchi}], it is indeed expected that its solution should depend on two integration `constants' -- actually functions of preferred time $T$. This matches precisely the result we obtained previously in spherical symmetry. While one of these functions of $T$ can be set to a desired value by the yet-to-be-fixed reparametrization freedom of $T$, Eq.~\eqref{T:reparam}, the second one will be related to the instantaneous mode of the theory [analogous to the function $\fIM(T)$ introduced above] and cannot be done away with even after fixing the said reparametrization freedom. 

The above logic does not apply to Einstein-{\ae}ther theory, simply because the {\ae}ther equation is not dynamically redundant even when the {\ae}ther is hypersurface orthogonal. Indeed, solutions of Eq.~\eqref{Ceqn:E:rho}, which obviously also hold in Einstein-{\ae}ther theory, do not always satisfy the {\ae}ther's equation of motion~\eqref{ae:EOM:ae}. 

Though slightly less rigorous than our spherically symmetric treatment, this last analysis has two advantages: it is more general and it clearly demonstrates that in Ho\v{r}ava gravity and in the preferred foliation the equations can be thought of as a system of an elliptic equation that needs to be imposed on every slice, elliptic equations that are preserved by time evolution and hence constitute constraints, and dynamical equations that generate the spacetime together with its foliation.  Reference~\cite{Donnelly:2011df} has reached the same conclusion by means of a  Hamiltonian analysis.

\section{Comments on spherical collapse}\label{collapse}
The problem of spherically symmetric collapse provides one of the simplest settings to which the preceding analysis can be directly applied, thereby allowing us to compare the dynamics of the two theories explicitly. Indeed, spherically symmetric collapse in Einstein-{\ae}ther theory have been previously studied in~\cite{Garfinkle:2007bk}, while an analogous simulation in Ho\v{r}ava theory is yet to be performed.\footnote{See, however, Ref.~\cite{Saravani:2013kva} where cuscuton theory is used as a proxy for Ho\v{r}ava gravity. We will discuss the relation between the two theories elsewhere.} In light of the relation between evolution in Einstein-{\ae}ther theory and Ho\v{r}ava gravity as discussed in the previous sections, it is tempting to revisit the results of Ref.~\cite{Garfinkle:2007bk}, potentially reinterpreting some of them, and to attempt to draw some general conclusions about spherically symmetric collapse in Ho\v{r}ava gravity.

To be more specific, in Ref.~\cite{Garfinkle:2007bk} spherically symmetric collapse in Einstein-{\ae}ther theory with a minimally coupled scalar field $\psi$ was studied, where $\psi$ represented collapsing matter. The evolution of the system was performed by adapting Eqs.~\eqref{ae:EOM} to the foliation described by the {\ae}ther that was hypersurface orthogonal due to spherical symmetry. This is the preferred foliation of Ho\v{r}ava gravity, as pointed out earlier. Equations.~\eqref{ae:EOM} were supplemented with appropriate equations of motion for $\psi$. 

Simulations were performed for two different values for the speed of the spin-$0$ mode $s_0$. In the first case the couplings $c_3$ and $c_4$ were set to zero, and the remaining two parameters of the theory, $c_1$ and $c_2$, were chosen such that $s_0$ was set to unity, \ie equal to the speed of light. Two values of $c_1$ were considered. For $c_1=0.7$ a regular (Killing) horizon forms as a result of the collapse while for $c_1=0.8$ no such horizon seems to form  and `$\dots$ the evolution seems to become singular, thus indicating the formation of a naked singularity.' The main reason for considering the specific values of the $c_i$ parameters and $s_0$ is because no static solutions had been found for the same values and $c_1\geq 0.8$ in Ref.~\cite{Eling:2006ec}. Indeed, the result was interpreted as verifying the absence of black holes for these parameters. However, static black holes were later found for that very same choice of the couplings in Ref.~\cite{Barausse:2011pu}, and it was argued there that the reason these solutions were not found in Ref.~\cite{Eling:2006ec} was insufficient accuracy in the numerics performed there. This puzzling situation definitely deserves further investigation. However, these simulations are not presented in detail in Ref.~\cite{Garfinkle:2007bk}, and so it is hard to interpret them in light of the later results of Ref.~\cite{Barausse:2011pu} or our analysis in the previous sections. Hence, we will not consider them further.

The second set of parameters was chosen such that the speed of the spin-$0$ mode was set to $\sqrt{2}$. With suitably chosen initial conditions, evolution led to the formation of a~\emph{regular} spin-$0$ horizon~\emph{inside} the metric horizon. Furthermore, at sufficiently `late times,' the geometry~\emph{outside the spin-$0$ horizon} settled down to the static solutions of~\cite{Eling:2006ec} to high accuracy.\ft{Note that~\cite{Garfinkle:2007bk} predates~\cite{Barausse:2011pu}, and thus were only able to compare their results with~\cite{Eling:2006ec}.} Moreover, the simulations of~\cite{Garfinkle:2007bk} also revealed that the  preferred frame lapse function $N$ `is driven to zero as the singularity is approached.' 

A vanishing of the lapse function at any given point of an evolution simulation in a gravity theory is strongly indicative of a breakdown of the corresponding foliation. A well known example of this is the study of spherically symmetric collapse in general relativity in  Schwarzschild coordinates, where a similar situation is expected toward the formation of the Killing horizon. On the other hand, provided one can be certain about the horizon-crossing properties of a certain foliation, having the lapse vanish asymptotically in time and as the singularity is approached is clearly advantageous from a numerical perspective. Since studying evolution with respect to the foliation defined by the {\ae}ther is merely a choice in Einstein-{\ae}ther theory, determining whether this is the optimal choice is a point that deserves further discussion.

The {\ae}ther's foliation in spherical symmetry will penetrate all Killing horizons, as the latter are null surfaces and the aether is always timelike. Considering also its privileged status in Einstein-aether theory, it was certainly a natural choice for Ref.~\cite{Garfinkle:2007bk}. One of the goals of Ref.~\cite{Garfinkle:2007bk} was indeed to verify whether regular spin-$0$ horizons emerge from spherical collapse in Einstein-{\ae}ther theory. Nonetheless, this foliation is special, and there is a way in which using it in this setting resembles using Schwarzschild coordinates in spherically symmetric collapse in general relativity: it does not penetrate the universal horizon.

Indeed, the vanishing of the lapse function $N$ in the preferred foliation can have an alternative interpretation as an asymptotic formation of a \emph{universal horizon}. In a static and spherically symmetric geometry, a universal horizon~\cite{Barausse:2011pu, Blas:2011ni} is a leaf of the preferred foliation that is also a constant $r$ hypersurface (and hence a hypersurface generated by the Killing vector associated with staticity), turning it into an event horizon even for arbitrarily fast propagations~\cite{Bhattacharyya:2015gwa}. In particular, the fact that a universal horizon is generated by a Killing vector implies~\cite{Bhattacharyya:2015gwa} that the preferred frame lapse function, subjected to the boundary condition~\eqref{N:bc}, will also vanish on the universal horizon. Moreover, a universal horizon can only occur in the asymptotic future in the preferred time. These observations, along with the fact that the geometry `outside' settles down to the appropriate static (and essentially unique~\cite{Blas:2011ni,Bhattacharyya:2014kta}) solution~\cite{Eling:2006ec, Barausse:2011pu}, thus strongly suggest that the simulations of Ref.~\cite{Garfinkle:2007bk} revealed the asymptotic formation of a universal horizon in the `late time' phase. Clearly, the notion of a universal horizon was introduced several years after Ref.~\cite{Garfinkle:2007bk} appeared, and it is natural that the above interpretation escaped its authors.

In situations where a universal horizon may form working in the preferred foliation is clearly not the optimum choice. The simulation will inevitably `stop' as the universal horizon is approached, and one may never cross it in this setup. If the simulations of Ref.~\cite{Garfinkle:2007bk} were to be performed again in a different foliation, it seems likely that one would be able to trace the formation and evolution of the universal horizon and verify whether the result leads to the static solutions of Ref.~\cite{Barausse:2011pu} all the way to the universal horizon and beyond. 

We now turn our attention to what the simulation of Ref.~\cite{Garfinkle:2007bk} can teach us about spherical collapse in Ho\v{r}ava gravity. Taking into account the connection between Ho\v{r}ava gravity and Einstein-{\ae}ther theory as discussed in detail in the previous sections, spherical collapse in the latter will be identical to spherical collapse in the former once  boundary conditions that set $\fIM(T)=0$ in~\eqref{ae:EOM:HL} have been chosen. The suitable boundary condition is simply regularity at the origin, $r=0$ (up to the formation of the singularity and/or universal horizon). Note that using the preferred foliation is not a choice but a necessity in Ho\v{r}ava gravity. Hence, the fact that the evolution seemingly `ends' with an asymptotic formation of a universal horizon appears to be a confirmation of the claim that the universal horizon is also a Cauchy horizon in theories like Ho\v{r}ava gravity~\cite{Blas:2011ni, Bhattacharyya:2015gwa}, where the  preferred foliation actually determines the causal structure and boundary data are required to determine the evolution.

\section{Conclusions}\label{conc}
Einstein-{\ae}ther theory with the additional constraint that the {\ae}ther be hypersurface orthogonal at the level of the field equations resembles the low-energy limit of Ho\v{r}ava gravity, but is a different theory. In both theories there is a special foliation but it has a different standing in each of them. This has been demonstrated clearly by comparing the initial value formulation in the two cases. In Ho\v{r}ava gravity the field equations are second order in time derivatives only in this foliation. Additionally, the system of evolution equations includes an elliptic equation that is not a constraint but instead needs to be imposed on each slice. The presence of such an elliptic equation implies that, in principle, evolution depends not only on initial data but also on future boundary/asymptotic data (for any type of boundary, including a conformal one).  This is a key feature of the causal structure of the theory~\cite{Bhattacharyya:2015gwa} and is intimately related with the presence of an instantaneous mode at the perturbative level~\cite{Blas:2011ni}. In other words, this foliation is both dynamically and causally preferred. In Einstein-{\ae}ther theory in contrast, the special foliation defined by the {\ae}ther is not preferred in any of these two senses. It does not define causality, and one need not adopt it to set up an initial value problem.

On the contrary, we have argued that choosing this foliation is not ideal when performing spherical collapse simulations in Einstein-{\ae}ther theory. This is because a universal horizon is actually a leaf of this foliation and the simulation cannot proceed past it in this slicing. This is reminiscent of spherical collapse in general relativity if performed in a foliation by constant Schwarzschild time surfaces. The collapse simulations of Ref.~\cite{Garfinkle:2007bk} have indeed been performed in the foliation defined by the hypersurface orthogonal {\ae}ther. We have revisited them and argued that they might have indeed uncovered the dynamical formation of a universal horizon asymptotically in time. In stationary black holes the lapse of this foliation vanishes on the universal horizon (when appropriately normalised at spatial infinity)~\cite{Bhattacharyya:2015gwa}, and in some of the simulations of Ref.~\cite{Garfinkle:2007bk}  the lapse indeed appears to vanish asymptotically in time. 

 Our results suggest that it would be particularly interesting to perform spherical collapse simulation in Einstein-{\ae}ther theory in a different foliation than that used in Ref.~\cite{Garfinkle:2007bk}. Such simulations would also effectively describe collapse in (low energy) Ho\v{r}ava gravity with additional regularity conditions at the center/universal horizon that determine the purely elliptic part of the evolution problem. Hence, they could shed light in the dynamical formation and evolution of universal horizons.

\acknowledgments

We would like to thank Enrico Barausse, Diego Blas, David Garfinkle, Ted Jacobson, Sergey Sibiryakov, and Helvi Witek for enlightening discussions.
The research leading to these results has received funding  from the European Research Council under the European Union's Seventh Framework Programme (FP7/2007-2013) / ERC Grant Agreement No. 306425 ``Challenging General Relativity.'' 
\appendix*
\section{Diffeomorphism invariance and the scalar equation}

Consider the Lagrangian density $\lag_1[\met^{a b}, \phi]$ and $\lag_2[\met^{a b}, u^d]$ that is a functional of the metric $\met^{a b}$ and the scalar field $\phi$.
The action of the diffeomorphism $\xi^b$ on $\lag_1$ can be written as
	\beql{L1}
	\begin{split}
	\LieD_{\xi}\lag_1[\met^{a b}, \phi] & = \frac{\delta \lag_1}{\delta \met^{a b}}\LieD_{\xi}\met^{a b} + \frac{\delta \lag_1}{\delta \phi}\LieD_{\xi} \phi \\
	                                    & \approx 2\left(\nabla^a \frac{\delta \lag_1}{\delta \met^{a b}}\right) \xi^b + \frac{\delta \lag_1}{\delta \phi}\xi^b\nabla_b \phi~,
	\end{split}
	\eeq
where $\approx$ is used to denote equality up to total divergences, which we are willing to neglect here. Note that $\delta \lag_1/\delta \met^{a b} = 0$ and $\delta \lag_1/\delta \phi = 0$ are by definition the field equations of the metric and the scalar, respectively. Hence, when the Lagrangian densities are invariant under diffeomorphisms the field equation for the scalar follows from the divergence of the field equation of the metric and vice versa, provided that $\phi$ is not constant.

The same calculation yields a different result when $\phi$ is replaced by a field with a higher tensorial rank. This is because the Lie derivative does not reduce to a directional derivative along the generator, as is the case for a scalar field. We will not repeat the calculation here, as it is essentially the same calculation that leads to Eqs.~(\ref{Bianchi}). For higher rank tensors, {\em e.g.}~vectors, having the field on shell implies that $\delta \lag_1/\delta \met^{a b}$ is divergence-free but the converse is not true.


\begin{thebibliography}{30}
%
\bibitem{Jacobson:2000xp} 
T.~Jacobson and D.~Mattingly, 
``Gravity with a dynamical preferred frame,'' 
Phys.\ Rev.\ D {\bf 64}, 024028 (2001) 
[gr-qc/0007031].
%
\bibitem{Horava:2009uw} 
P.~Horava, 
``Quantum Gravity at a Lifshitz Point,'' 
Phys.\ Rev.\ D {\bf 79}, 084008 (2009) 
[arXiv:0901.3775 [hep-th]].
%
\bibitem{Elliott:2005va} 
J.~W.~Elliott, G.~D.~Moore and H.~Stoica, 
``Constraining the new {\ae}ther: Gravitational \v{C}erenkov radiation,'' 
JHEP {\bf 0508}, 066 (2005) 
[hep-ph/0505211].
%
\bibitem{Eling:2006ec} 
C.~Eling and T.~Jacobson, 
``Black Holes in Einstein-{\ae}ther Theory,'' 
Class.\ Quant.\ Grav.\  {\bf 23}, 5643 (2006) [Class.\ Quant.\ Grav.\  {\bf 27}, 049802 (2010)] 
[gr-qc/0604088].
%
\bibitem{Blas:2011ni} 
D.~Blas and S.~Sibiryakov, 
``Ho\v{r}ava gravity versus thermodynamics: The Black hole case,'' 
Phys.\ Rev.\ D {\bf 84}, 124043 (2011) 
[arXiv:1110.2195 [hep-th]].
%
\bibitem{Barausse:2011pu} 
E.~Barausse, T.~Jacobson and T.~P.~Sotiriou, 
``Black holes in Einstein-{\ae}ther and Ho\v{r}ava-Lifshitz gravity,''
Phys.\ Rev.\ D {\bf 83}, 124043 (2011) 
[arXiv:1104.2889 [gr-qc]].
%
\bibitem{Bhattacharyya:2015gwa} 
J.~Bhattacharyya, M.~Colombo and T.~P.~Sotiriou, 
``Causality and black holes in spacetimes with a preferred foliation,'' 
[arXiv:1509.01558 [gr-qc]].
%
\bibitem{Berglund:2012bu} 
  P.~Berglund, J.~Bhattacharyya and D.~Mattingly,
  ``Mechanics of universal horizons,''
  Phys.\ Rev.\ D {\bf 85}, 124019 (2012)
  [arXiv:1202.4497 [hep-th]].
%
\bibitem{Barausse:2012qh} 
  E.~Barausse and T.~P.~Sotiriou,
  ``Slowly rotating black holes in Ho\v{r}ava-Lifshitz gravity,''
  Phys.\ Rev.\ D {\bf 87}, 087504 (2013)
  [arXiv:1212.1334].
%
\bibitem{Sotiriou:2014gna} 
  T.~P.~Sotiriou, I.~Vega and D.~Vernieri,
  ``Rotating black holes in three-dimensional Ho\v{r}ava gravity,''
  Phys.\ Rev.\ D {\bf 90}, no. 4, 044046 (2014)
  [arXiv:1405.3715 [gr-qc]].
%
\bibitem{Jacobson:2010mx} 
T.~Jacobson, 
``Extended Ho\v{r}ava gravity and Einstein-{\ae}ther theory,'' 
Phys.\ Rev.\ D {\bf 81}, 101502 (2010) [Phys.\ Rev.\ D {\bf 82}, 129901 (2010)] 
[arXiv:1001.4823 [hep-th]].
%
\bibitem{Barausse:2012ny} 
  E.~Barausse and T.~P.~Sotiriou,
  ``A no-go theorem for slowly rotating black holes in Ho\v{r}ava-Lifshitz gravity,''
  Phys.\ Rev.\ Lett.\  {\bf 109}, 181101 (2012) [Phys.\ Rev.\ Lett.\  {\bf 110}, no. 3, 039902 (2013)]
  [arXiv:1207.6370].
%
\bibitem{Garfinkle:2007bk} 
D.~Garfinkle, C.~Eling and T.~Jacobson, 
``Numerical simulations of gravitational collapse in Einstein-{\ae}ther theory,'' 
Phys.\ Rev.\ D {\bf 76}, 024003 (2007) 
[gr-qc/0703093 [GR-QC]].
%
\bibitem{Saravani:2013kva} 
  M.~Saravani, N.~Afshordi and R.~B.~Mann,
  ``Dynamical Emergence of Universal Horizons during the formation of Black Holes,''
  Phys.\ Rev.\ D {\bf 89}, no. 8, 084029 (2014)
  [arXiv:1310.4143 [gr-qc]].
%
\bibitem{Blas:2009qj} 
  D.~Blas, O.~Pujolas and S.~Sibiryakov,
  ``Consistent Extension of Ho\v{r}ava Gravity,''
  Phys.\ Rev.\ Lett.\  {\bf 104}, 181302 (2010)
  [arXiv:0909.3525 [hep-th]].
%
\bibitem{Sotiriou:2009gy} 
  T.~P.~Sotiriou, M.~Visser and S.~Weinfurtner,
  ``Phenomenologically viable Lorentz-violating quantum gravity,''
  Phys.\ Rev.\ Lett.\  {\bf 102}, 251601 (2009)
  [arXiv:0904.4464 [hep-th]].
%
\bibitem{Sotiriou:2009bx} 
  T.~P.~Sotiriou, M.~Visser and S.~Weinfurtner,
  ``Quantum gravity without Lorentz invariance,''
  JHEP {\bf 0910}, 033 (2009)
  [arXiv:0905.2798 [hep-th]].
%
\bibitem{Weinfurtner:2010hz} 
  S.~Weinfurtner, T.~P.~Sotiriou and M.~Visser,
  ``Projectable Ho\v{r}ava-Lifshitz gravity in a nutshell,''
  J.\ Phys.\ Conf.\ Ser.\  {\bf 222}, 012054 (2010)
  [arXiv:1002.0308 [gr-qc]].
%
\bibitem{Mukohyama:2010xz} 
  S.~Mukohyama,
  ``Ho\v{r}ava-Lifshitz Cosmology: A Review,''
  Class.\ Quant.\ Grav.\  {\bf 27}, 223101 (2010)
  [arXiv:1007.5199 [hep-th]].
%
\bibitem{Vernieri:2011aa} 
  D.~Vernieri and T.~P.~Sotiriou,
  ``Ho\v{r}ava-Lifshitz Gravity: Detailed Balance Revisited,''
  Phys.\ Rev.\ D {\bf 85}, 064003 (2012)
  [arXiv:1112.3385 [hep-th]].
%
\bibitem{Vernieri:2012ms} 
  D.~Vernieri and T.~P.~Sotiriou,
  ``Ho\v{r}ava-Lifshitz gravity with detailed balance,''
  J.\ Phys.\ Conf.\ Ser.\  {\bf 453}, 012022 (2013)
  [arXiv:1212.4402 [hep-th]].
%
\bibitem{Sotiriou:2010wn} 
  T.~P.~Sotiriou,
  ``Ho\v{r}ava-Lifshitz gravity: a status report,''
  J.\ Phys.\ Conf.\ Ser.\  {\bf 283}, 012034 (2011)
  [arXiv:1010.3218 [hep-th]].
%
\bibitem{Sotiriou:2011dr} 
  T.~P.~Sotiriou, M.~Visser and S.~Weinfurtner,
  ``Lower-dimensional Ho\v{r}ava-Lifshitz gravity,''
  Phys.\ Rev.\ D {\bf 83}, 124021 (2011)
  [arXiv:1103.3013 [hep-th]].
%
\bibitem{Blas:2010hb} 
  D.~Blas, O.~Pujolas and S.~Sibiryakov,
  ``Models of non-relativistic quantum gravity: The Good, the bad and the healthy,''
  JHEP {\bf 1104}, 018 (2011)
  [arXiv:1007.3503 [hep-th]].
%
\bibitem{Bhattacharyya:2014kta} 
J.~Bhattacharyya and D.~Mattingly, 
``Universal horizons in maximally symmetric spaces,'' 
Int.\ J.\ Mod.\ Phys.\ D {\bf 23}, no. 13, 1443005 (2014) 
[arXiv:1408.6479 [hep-th]].
%
\bibitem{Wald:1984rg} 
R.~M.~Wald, 
``General Relativity"  The University of Chicago Press, Chicago 1984.
%
\bibitem{Jacobson:2011cc} 
T.~Jacobson, 
``Initial value constraints with tensor matter,'' 
Class.\ Quant.\ Grav.\  {\bf 28}, 245011 (2011) 
[arXiv:1108.1496 [gr-qc]].
%
\bibitem{Donnelly:2011df} 
W.~Donnelly and T.~Jacobson, 
``Hamiltonian structure of Ho\v{r}ava gravity,'' 
Phys.\ Rev.\ D {\bf 84}, 104019 (2011) 
[arXiv:1106.2131 [hep-th]].
%
\end{thebibliography}
\end{document}